\documentclass[modern]{aastex61}

\usepackage{graphicx}
\usepackage{amsmath}
\usepackage{amssymb}
\usepackage{enumitem}

\newcommand\mybar{\kern1pt\rule[-\dp\strutbox]{.8pt}{\baselineskip}\kern1pt}

\setlist[itemize]{noitemsep, topsep=0pt, leftmargin=*}

\shorttitle{LRDs from Low-Spin Galaxies}
\shortauthors{Loeb}

\begin{document}

\title{Little Red Dots from Low-Spin Galaxies at High Redshifts}

\author{Abraham Loeb}
\affiliation{Astronomy Department, Harvard University, 60 Garden
  St., Cambridge, MA 02138, USA}

\begin{abstract}
  Recently, a new population of compact, high-redshift ($z\gtrsim 7$)
  galaxies appeared as little red dots (LRDs) in deep JWST
  observations. The latest spectroscopic data indicates that these
  galaxies contain an evolved stellar population, reflecting an early
  episode of high star-formation-rate. The appearance of broad
  emission lines suggests that a central overmassive black hole also
  powers these galaxies. I propose that LRD galaxies represent the
  low-spin tail of the galaxy population. Low-spin galaxies host a
  more compact gaseous disk with an enhanced star formation rate
  relative to typical galaxies at the same redshift. The compact disk
  feeds efficiently a central black hole, as predicted by
  ~\citet{Eisen96}.
\end{abstract}

\section{Introduction}

One of the surprising discoveries of the James Webb Space Telescope
(JWST) involves an early population of compact red galaxies at
redshifts $z\gtrsim 7$, when the Universe was $\lesssim 800$~Myr
old~\citep{Labbe23}. These galaxies, commonly dubbed as little red
dots (LRDs), are redder than expected from their cosmological
redshift, indicating additional reddening by dust. Spectroscopy of
some LRD galaxies indicates that they may contain as much mass in
evolved stars as the Milky Way galaxy or even more, up to $\sim
10^{11}M_\odot$~\citep{Wang24}. Nevertheless, LRDs have an effective
radius $\sim 10^2$pc, smaller by a factor of 10-100 relative to
expectations.

The mass in stars needed to light up LRDs requires, in the context of
the expected LCDM abundance of galactic halos, that they convert
nearly all their gas into stars rapidly, over hundreds of millions of
years~\citep{Boylan23,Wang24}. This high level of star formation
efficiency is unlikely to be realized because of the inevitable gas
loss in supernova-driven winds~\citep{LF12}. This suggests that a
significant fraction of the light emitted by LRDs is powered by a
central overmassive black hole~\citep{Pacucci23,Pacucci24}.

The existence of a black hole in LRDs is supported by the
spectroscopic detection of broad emission lines, indicating motions of
line-emitting gas with a speed of up to $\sim 2.5\times 10^3~{\rm
  km~s^{-1}}$~\citep{Wang24}, as expected from the broad-line-region
in the vicinity of a black hole. So far, no X-rays were
detected from LRDs~\citep{Ananna24,Yue24}.  The required mass of the
black hole is above expectations based on the stellar mass-black hole
mass correlation in the present-day universe~\citep{Wang24,Pacucci23}.

Here, I suggest that LRDs are drawn from the low-spin tail in the
distribution of specific angular momenta of
galaxies~\citep{Eisen95,Eisen96}. Such an origin would account for
their compact disk, high star-formation-rate, enhanced dust opacity,
and overmassive black hole.

\section{Low-Spin Galaxies}

The standard semi-analytic model for galactic disks relates the disk
radius to the angular momentum per unit mass of the host halo,
$j$~\citep{MoMau}. Galactic spin is acquired via tidal torques during
turnaround of the protogalactic
material~\citep{Peebles,Eisen95,White_Loeb}. As the dark matter
virializes, the gas cools and settles to a disk of mass $M_d$. The
disk radius is dictated by the centrifugal barrier, $r_d\approx
j^2/GM_d$. Since different galaxies form in different environments,
the population of disks will have a distribution of $j$-values at
birth. Figure 1 in~\citet{Eisen95} suggests that the number density of
galactic disks which are 10 to 100 times smaller than a typical radius
$\langle r_d \rangle$ is $\sim 10\%$ to $\sim 1\%$, respectively, of
the abundance of typical galaxies with the same mass at the same
redshift. Low-spin galaxies could therefore account for a population
of LRD disks with the required abundance and compactness to match the
JWST data~\citep{Wang24}.

According to the Kennicutt-Schmidt prescription~\citep{Ken}, the star
formation rate scales inversely with the dynamical time of the disk,
$t_{\rm dyn}\sim 1/\sqrt{GM_d/r_d^{3}}$. The reduction in $r_d$ for LRDs
naturally leads to an enhanced star formation rate, ${\rm SFR}\propto
(M_d/t_{\rm dyn})\sim \sqrt{GM_d^3/r_d^3}$. A reduction in $r_d$ by a
factor of 10 to 100 leads to an increased SFR by factors of $\sim 32$
to $10^3$, respectively. This could account for an early episode of
high SFR and the resulting evolved stellar population in LRDs. The
early SFR would naturally enrich the disk with dust that together with
the disk compactness, would increase the dust opacity and result in
enhanced redenning as observed for LRDs.

\section{Black Hole Formation}

A reduction in $r_d$ relative to typical values by factors of 10-100
leads to a decrease in the centrifugal barrier for feeding a central
black hole.  Consider a halo mass of $M_h\sim 10^{12}M_\odot$, and a
disk mass containing a third of the total baryonic mass in the halo,
$M_d \approx 0.05M_h=5\times 10^{10}M_\odot$. A disk smaller by a
factor $\sim 25$ relative to the average value $\langle r_d\rangle$
for the population of galaxies with the same mass and
redshift~\citep{Eisen96}, would have a radius comparable to LRDs,
\begin{equation}
r_d \approx 100~{\rm pc} \left({r_d\over 0.04 \langle
  r_d\rangle}\right)\times \left({M_d\over 5\times
  10^{10}M_\odot}\right)^{0.4} .
\label{eqone}
\end{equation}
The characteristic circular velocity of a disk of this mass and
radius, $v_c=(GM_d/r_d)^{1/2}\approx 1.5\times 10^3~{\rm km~s^{-1}}$,
is comparable - as needed - to half the velocity width of the observed
broad lines of LRDs~\citep{Wang24}.

Based on Figure 3 in~\citet{Eisen96}, the required comoving density of
LRD halos, $\gtrsim 10^{-5}~{\rm Mpc^{-3}}$, allows for overmassive
black hole progenitors with masses $\lesssim 5\times 10^8M_\odot$ and
viscous evolution times of their progenitor disks $\lesssim 10^8~{\rm
  yr}$, as needed for LRDs~(see Figure 8 in~\citet{Wang24}) .

\bigskip
\bigskip
\bigskip
\bigskip
\section*{Acknowledgements}

I thank Fabio Pacucci for inspiring this work. This research was
supported in part by Harvard's {\it Black Hole Initiative}, which is
funded by grants from JFT and GBMF.

\bigskip
\bigskip
\bigskip

\bibliographystyle{aasjournal}
\bibliography{t}

\begin{thebibliography}{}
\expandafter\ifx\csname natexlab\endcsname\relax\def\natexlab#1{#1}\fi
\providecommand{\url}[1]{\href{#1}{#1}}
\providecommand{\dodoi}[1]{doi:~\href{http://doi.org/#1}{\nolinkurl{#1}}}
\providecommand{\doeprint}[1]{\href{http://ascl.net/#1}{\nolinkurl{http://ascl.net/#1}}}
\providecommand{\doarXiv}[1]{\href{https://arxiv.org/abs/#1}{\nolinkurl{https://arxiv.org/abs/#1}}}

\bibitem[{{Ananna} {et~al.}(2024){Ananna}, {Bogd{\'a}n}, {Kov{\'a}cs},
  {Natarajan}, \& {Hickox}}]{Ananna24}
{Ananna}, T.~T., {Bogd{\'a}n}, {\'A}., {Kov{\'a}cs}, O.~E., {Natarajan}, P., \&
  {Hickox}, R.~C. 2024, arXiv e-prints, arXiv:2404.19010,
  \dodoi{10.48550/arXiv.2404.19010}

\bibitem[{{Boylan-Kolchin}(2023)}]{Boylan23}
{Boylan-Kolchin}, M. 2023, Nature Astronomy, 7, 731,
  \dodoi{10.1038/s41550-023-01937-7}

\bibitem[{{Eisenstein} \& {Loeb}(1995{\natexlab{a}})}]{Eisen96}
{Eisenstein}, D.~J., \& {Loeb}, A. 1995{\natexlab{a}}, \apj, 443, 11,
  \dodoi{10.1086/175498}

\bibitem[{{Eisenstein} \& {Loeb}(1995{\natexlab{b}})}]{Eisen95}
---. 1995{\natexlab{b}}, \apj, 439, 520, \dodoi{10.1086/175193}

\bibitem[{{Gao} {et~al.}(2004){Gao}, {Loeb}, {Peebles}, {White}, \&
  {Jenkins}}]{White_Loeb}
{Gao}, L., {Loeb}, A., {Peebles}, P.~J.~E., {White}, S. D.~M., \& {Jenkins}, A.
  2004, \apj, 614, 17, \dodoi{10.1086/423444}

\bibitem[{{Kennicutt} \& {Evans}(2012)}]{Ken}
{Kennicutt}, R.~C., \& {Evans}, N.~J. 2012, \araa, 50, 531,
  \dodoi{10.1146/annurev-astro-081811-125610}

\bibitem[{{Labb{\'e}} {et~al.}(2023){Labb{\'e}}, {van Dokkum}, {Nelson},
  {Bezanson}, {Suess}, {Leja}, {Brammer}, {Whitaker}, {Mathews}, {Stefanon}, \&
  {Wang}}]{Labbe23}
{Labb{\'e}}, I., {van Dokkum}, P., {Nelson}, E., {et~al.} 2023, \nat, 616, 266,
  \dodoi{10.1038/s41586-023-05786-2}

\bibitem[{{Loeb} \& {Furlanetto}(2013)}]{LF12}
{Loeb}, A., \& {Furlanetto}, S.~R. 2013, {The First Galaxies in the Universe}
  ({Princeton University Press})

\bibitem[{{Mo} {et~al.}(1998){Mo}, {Mao}, \& {White}}]{MoMau}
{Mo}, H.~J., {Mao}, S., \& {White}, S. D.~M. 1998, \mnras, 295, 319,
  \dodoi{10.1046/j.1365-8711.1998.01227.x}

\bibitem[{{Pacucci} \& {Loeb}(2024)}]{Pacucci24}
{Pacucci}, F., \& {Loeb}, A. 2024, \apj, 964, 154,
  \dodoi{10.3847/1538-4357/ad3044}

\bibitem[{{Pacucci} {et~al.}(2023){Pacucci}, {Nguyen}, {Carniani}, {Maiolino},
  \& {Fan}}]{Pacucci23}
{Pacucci}, F., {Nguyen}, B., {Carniani}, S., {Maiolino}, R., \& {Fan}, X. 2023,
  \apjl, 957, L3, \dodoi{10.3847/2041-8213/ad0158}

\bibitem[{{Peebles}(1969)}]{Peebles}
{Peebles}, P.~J.~E. 1969, \apj, 155, 393, \dodoi{10.1086/149876}

\bibitem[{{Wang} {et~al.}(2024){Wang}, {Leja}, {de Graaff}, {Brammer},
  {Weibel}, {van Dokkum}, {Baggen}, {Suess}, {Greene}, {Bezanson}, {Cleri},
  {Hirschmann}, {Labb{\'e}}, {Matthee}, {McConachie}, {Naidu}, {Nelson},
  {Oesch}, {Setton}, \& {Williams}}]{Wang24}
{Wang}, B., {Leja}, J., {de Graaff}, A., {et~al.} 2024, \apjl, 969, L13,
  \dodoi{10.3847/2041-8213/ad55f7}

\bibitem[{{Yue} {et~al.}(2024){Yue}, {Eilers}, {Ananna}, {Panagiotou}, {Kara},
  \& {Miyaji}}]{Yue24}
{Yue}, M., {Eilers}, A.-C., {Ananna}, T.~T., {et~al.} 2024, arXiv e-prints,
  arXiv:2404.13290, \dodoi{10.48550/arXiv.2404.13290}

\end{thebibliography}
\label{lastpage}
\end{document}